\documentclass[showpacs,twocolumn,showkeys,amsmath,amssymb,pra]{revtex4-1}
\usepackage{graphicx}
\usepackage{bm} 
        
\begin{document}

\title[Two-particle NOON-states in periodically shaken three-well potentials]{Spatial two-particle NOON-states  in periodically shaken three-well potentials}

\author{Kirsten Stiebler}
\author{Bettina Gertjerenken}
\author{Niklas Teichmann}
\author{Christoph Weiss}
\email{Christoph.Weiss@uni-oldenburg.de}

\affiliation{Institut f\"ur Physik, Carl von Ossietzky Universit\"at,
                D-26111 Oldenburg, Germany
}

\date{\today}
 
\begin{abstract}
Few-particle dynamics in a three-well potential are investigated numerically. It is shown that periodically shaking the potential can considerably increase the fidelity of emerging spatial quantum superpositions. Such NOON-states are important for quantum interferometry. If the two particles initially sit in the middle well, the probability to return to this state can distinguish pure quantum dynamics from statistical mixtures. The numeric implementation of decoherence via particle losses shows clear differences from the pure quantum behaviour. A three-well lattice could be an ideal system for experimental realisations.
\end{abstract} 
\maketitle 

\section{Introduction}

Double-well lattices~\cite{SebbyStrableyEtAl07,YukalovYukalova09} provide an interesting, experimentally accessible tool to investigate quantum effects for atom numbers even down to less than six atoms~\cite{CheinetEtAl08}. Three-well lattices and four-well lattices could be realised either by using subwavelength lattices~\cite{YiEtAl08} or by adjusting the technique used in Refs.~\cite{CheinetEtAl08,Folling07} by using more lasers. The number of particles in one of the two wells of a double-well lattice can easily be measured by averaging over all double wells~\cite{CheinetEtAl08}. For a single three-well potential, transport and interaction blockade of cold bosonic atoms have been discussed in Ref.~\cite{SchlagheckEtAl10}. In a single three-well potential, few particle dynamics could be observed by using the single-site addressability of Refs.~\cite{BakrEtAl09,ShersonEtAl10}.

One of the fascinating aspects of quantum mechanics are non-classical quantum states like the NOON-states~\cite{Wildfeuer07}:
\begin{equation}
|\Psi_{\rm NOON}\rangle \equiv \frac1{\sqrt{2}} \left(|N,0\rangle +|0,N\rangle\right)\;,
\end{equation}
where $|N-n,n\rangle$ refers, e.g., to $N-n$ particles being on the left and $n$ particles on the right of some barrier but NOON-states can also be investigated in phase-space. These states have also been called ``Schr\"odinger cat''-states, e.g., in Refs.~\cite{MonroeEtAl96,BruneEtAl96}. Like the spin-squeezed states of Refs.~\cite{SorensenEtAl01,EsteveEtAl08}, such states are relevant to improve interferometric measurements~\cite{GiovannettiEtAl04}.
Suggestions on how such interesting many-particle quantum superpositions might be
obtained can be found, e.g., in Refs.~\cite{CastinDalibard97,
RuostekoskiEtAl98, CiracEtAl98, DunninghamBurnett01, MicheliEtAl03, MahmudEtAl05,Dounas-frazerEtAl07,WeissCastin09,DagninoEtAl09,StreltsovEtAl09,GarciaMarchEtAl10b} and references therein.  

The focus of the present paper lies on the numeric creation of spatial two-particle NOON-states in three-well potentials with a focus on experimental signatures to verify that indeed NOON-states rather than statistical mixtures have been created. To achieve the NOON-states via controlled quantum dynamics, we suggest to
periodically shake the three-well potential. This approach to enhance tunnelling in a controlled way is based on photon-assisted tunnelling~\nocite{KohlerSols03}\nocite{EckardtEtAl05}\nocite{TeichmannEtAl09}\nocite{XieEtAl10}\cite{KohlerSols03,EckardtEtAl05,TeichmannEtAl09,XieEtAl10,EsmannEtAl10IOP} which was realised experimentally for Bose-Einstein condensates in optical lattices~\cite{SiasEtAl08} (for the interpretation of the experimental data see also \cite{CreffieldEtAl10}). Recent  research in periodically driven systems include nonlinear Landau-Zener processes~\cite{ZhangEtAl08} and transport of bound pairs~\cite{KudoEtAl09,WeissBreuer09}.

The paper is organised as follows: in Sec.~\ref{sec:Model} we introduce the Hamiltonian used to model the periodically shaken three-well potential as well as our main aim, the  two-particle NOON-states.  Section~\ref{sec:Simplified} shows that for some parameters, the quantum dynamics of the periodically shaken system can be understood within a simplified model originally developed for double wells~\cite{EsmannEtAl10IOP} which we adopt for the present situation. In Sec.~\ref{sec:How} we introduce the two-particle NOON-states and explain how they can be distinguished from statistical mixtures. Section~\ref{sec:NOON} shows that periodic shaking can considerably increase the fidelity, i.e., the quality of the quantum superpositions. In order to show that the quantum superpositions can indeed be distinguished from statistical mixtures, last but not least decoherence effects are included.

\section{\label{sec:Model}Model system}
\begin{figure}
\begin{center}
\includegraphics[width=0.6\linewidth]{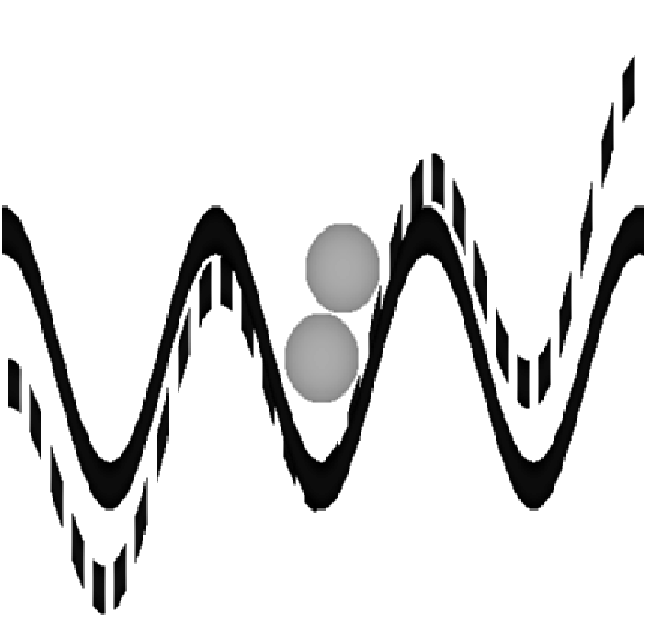}
\end{center}
\caption{\label{fig:sketch}
The three-well potential with initially two atoms in the middle well. This situation is an experimentally realistic initial condition~\cite{Schreck09}. Experimental investigations would, e.g., be possible in three-well lattices~\cite{YiEtAl08} (cf.\ the double-well lattice potential with less than six atoms per lattice site of Ref.~\cite{CheinetEtAl08}). Periodic shaking is currently established as an experimental tool to control the quantum dynamics~\cite{SiasEtAl08,CreffieldEtAl10,KierigEtAl08,EckardtEtAl09b}.
}
\end{figure}

For the description of ultracold atoms in a three-well potential, the three-site version of a many-particle Hamiltonian~\cite{MilburnEtAl97}  originally developed in nuclear physics~\cite{LipkinEtAl65} is used:  
\begin{eqnarray}
\label{eq:H}
 \hat{H}&=&-J\left(\hat{c}_1^{\dag}\hat{c}_2{\phantom\dag}+\hat{c}_2^{\dag}\hat{c}_1{\phantom\dag}+\hat{c}_3^{\dag}\hat{c}_2{\phantom\dag}+\hat{c}_2^{\dag}\hat{c}_3{\phantom\dag}\right)\nonumber\\
        &+&\frac{U}{2}\sum\limits_{i=1}^3\hat{c}_i^{\dag}\hat{c}_i^{\dag}\hat{c}_i^{\phantom\dag}\hat{c}_i^{\phantom\dag}                  \nonumber\\
        &+&2\left[\hbar\mu_0+\hbar\mu_1\sin(\omega t)\right]\left(\hat{c}_1^{\dag}\hat{c}_1^{\phantom\dag}-\hat{c}_3^{\dag}\hat{c}_3^{\phantom\dag}\right)
\end{eqnarray}
The operator $\hat{c}^{(\dag)}_j$ annihilates (creates) a boson in well~$j$;
$J$ is the hopping matrix element, $2\hbar\mu_0$ is
the tilt between well~1 and well~3 and $2\hbar\mu_1$ is the driving amplitude. The interaction
between a pair of particles in the same well is denoted by $U$. Quantum dynamics beyond such models has been investigated, e.g., by Refs.~\cite{SchlagheckEtAl10,TrimbornEtAl09,SakmannEtAl09,ZollnerEtAl08}~\footnote{{Reference~\cite{CheinetEtAl08} demonstrates that the Hamiltonian~(\ref{eq:H}) can indeed be valid to describe the dynamics of few particles in few well systems.}}. A systematic sketch of the three-well potential is shown in Fig.~\ref{fig:sketch}.

A Fock-basis is useful to describe $N$ atoms in the three-well potential:
\begin{eqnarray}
\label{eq:Fock}
|\psi_{N-n-m,n,m}\rangle&=&|N-n-m,n,m\rangle\nonumber\\
&\equiv & |n,m\rangle
\label{eq:Fochshort}
\end{eqnarray}
with $m$ particles in the right well, $n$ particles in the middle well and $N-m-n$ particles in the left well.

The aim of this paper is not to produce a single-particle quantum superposition but a two-particle NOON-state:
\begin{equation}
\label{eq:cat}
|\Psi_{\delta}\rangle = \frac1{\sqrt{2}}\left(|2,0,0\rangle+e^{i\delta}|0,0,2\rangle\right)\;.
\end{equation}
As will be shown in this paper, periodic shaking helps on the way to achieve this aim. In order to characterise the quality with which this aim is achieved for a given normalised wave-function $|\psi(t)\rangle$ ($\langle\psi(t)|\psi(t)\rangle=1$), the fidelity, 
\begin{equation}
F_{\delta} = \left|\langle\psi(t)|\Psi_{\delta}\rangle\right|^2\;,
\end{equation}
or more precise the fidelity maximised over all possible angles~$\delta$ is used:
\begin{equation}
\label{eq:fidelity}
{\rm fidelity} = {\rm max}\left(F_{\delta}, 0\le\delta< 2\pi\right).
\end{equation}
In order to be a useful flag to indicate a NOON-state, the fidelity has to be larger than $50\%$; below this value, the state might not be a superposition at all as, e.g., the state $|2,0,0\rangle$ would also result in a fidelity of $0.5$.

\begin{figure}
\begin{center}
\includegraphics[width = 1.0\linewidth]{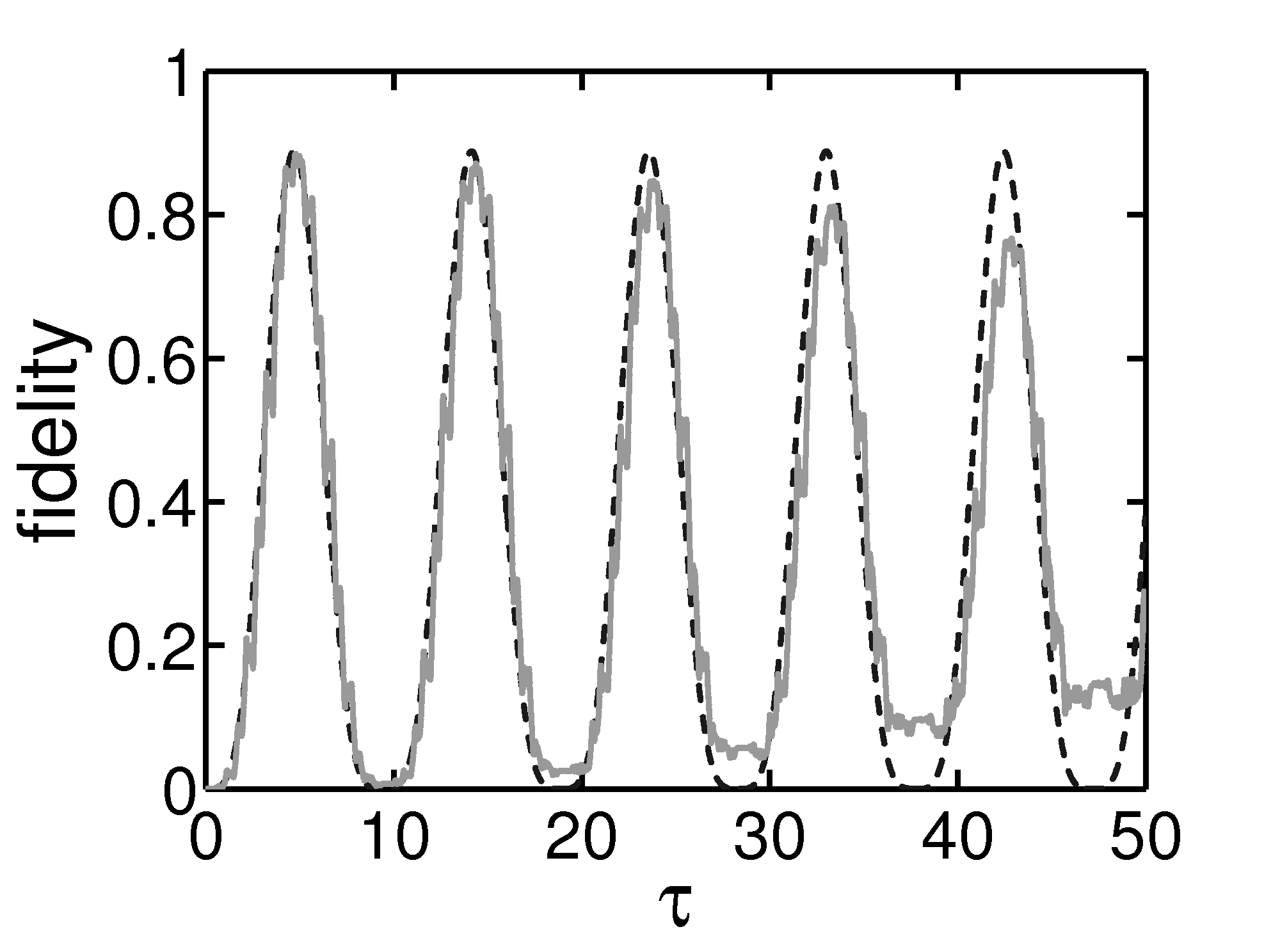}
\end{center}
\caption{\label{fig:exactdiagonal}Fidelity~(\ref{eq:fidelity}) as a function of dimensionless time $\tau$ [Eq.~(\ref{eq:tau})] for the experimentally realistic initial condition~\cite{Schreck09} of two particles sitting in the middle well of a three-well potential. The parameters were chosen such that shaking frequency, interaction and tilt are equal: $\hbar \omega=U=\hbar \mu_0=6J$ which corresponds to integer photon-assisted tunnelling~\cite{EckardtEtAl05,XieEtAl10}. The shaking amplitude was chosen to be $\hbar\mu_1=25.2J$. It has been optimised with the simplified model~(\ref{eq:simplified}). Solid grey line: the exact numerical solution of the time-dependent model Hamiltonian~(\ref{eq:H}). Dashed black line: the simplified model which can be solved via exact diagonalisation. While the fidelity is well above 50\%, it still remains below 90\% (the exact numerics reaches the maximum $0.8852$ at $\tau=4.787$).
}
\end{figure}
\begin{figure}
\begin{center}
\includegraphics[width = 1.0\linewidth]{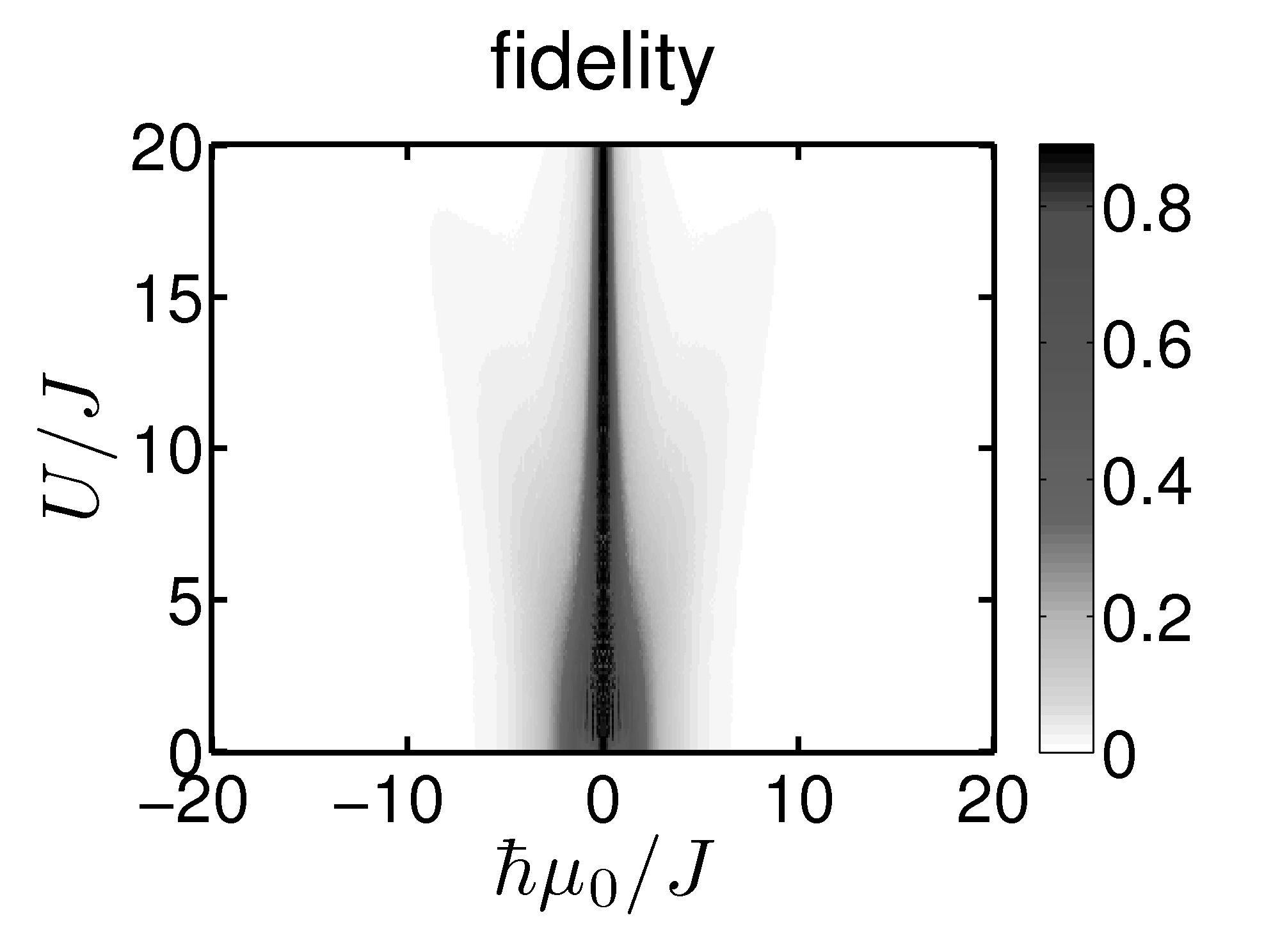}
\end{center}
\caption{\label{fig:ohnetreiben}Fidelity in a two-dimensional projection as a function of both interaction~$U$ and tilt $\hbar\mu_0$ without any shaking. To obtain the plotted data, the model~(\ref{eq:H}) was solved numerically for two particles initially in the middle well and times~$\tau$ with $0\le\tau\le 50$. for each set of parameters, the maximum is plotted. Similar to the periodically shaken case displayed in Fig.~\ref{fig:exactdiagonal}, the fidelity stays below 90\% ($0.892$ at $U=1.3J$, $\hbar \mu_0=-0.5J$ and $\tau\simeq 15$). Higher fidelities will be achieved via periodic shaking (cf.\ Fig.~\ref{fig:fidgut}) 
 }
\end{figure}

Figure~\ref{fig:exactdiagonal} displays an example for which the fidelity is well above $50\%$. The data is plotted as a function of the dimensionless time,
\begin{equation}
\label{eq:tau}
\tau \equiv \frac{Jt}{\hbar}\;.
\end{equation}
Figure~\ref{fig:ohnetreiben} shows that for the timescales of Fig.~\ref{fig:exactdiagonal} and a large parameter range, even without shaking, the fidelity can reach values well above~50\% (Fig.~\ref{fig:ohnetreiben}) but it stays again below 90\% (Fig.~\ref{fig:ohnetreiben}). Thus, so far there does not seem to be an advantage of using periodic shaking. However, before showing that periodic shaking can lead to fidelities well above 95\%, we show that it is possible to quantitatively understand the behaviour displayed in Fig.~\ref{fig:exactdiagonal}.

\section{\label{sec:Simplified}Simplified model}

 In order to understand the behaviour under periodic shaking, a simplified version of the Hamiltonian~(\ref{eq:H}) can be used. This approach introduced below will be useful for large shaking frequencies and not too long timescales~\cite{EsmannEtAl10IOP}. 

Extending the simplified model of Ref.~\cite{EsmannEtAl10IOP} to a three-well potential starts with writing the Schr\"odinger equation corresponding to the Hamiltonian~(\ref{eq:H}) in the Fock basis~(\ref{eq:Fock}). Using the short notation of Eq.~(\ref{eq:Fochshort}), we then introduce the ansatz motivated by the interaction picture
\begin{equation}
\label{eq:ansatz}
   \langle \nu,\mu|\psi(t)\rangle = 
   a_{\nu,\mu}(t)\exp\left[-\frac{i}{\hbar}\int
   \langle \nu,\mu | H_0(t)|\nu,\mu\rangle d t\right] \;,
\end{equation}
with $H_0 \equiv \lim_{J\to 0} H$. The tunnelling between Fock states is then included in a system of first-order one-dimensional differential equations for the $a_{\nu,\mu}$ (cf.\ Refs.~\cite{TeichmannEtAl09,EsmannEtAl10IOP}). The time-dependence enters entirely via the exponents introduced in the ansatz~(\ref{eq:ansatz}); it includes terms linear in time (either proportional to the interaction or to the tilt) as well as a term proportional to the driving amplitude for which the time-dependence is a $\cos(\omega t)$. This last term can be transferred into a sum of exponentials with linear time-dependence via~\cite{Abramowitz84}
\begin{equation}
\label{eq:bessel}
e^{iz\cos(\omega t)}=\sum_{k=-\infty}^{\infty}{\cal J}_{k}(z)i^ke^{ik\omega t}\,,
\end{equation}
where the ${\cal J}_{k}(z)$ are Bessel functions. This leads to a series of relevant frequencies which for two particles read:
\begin{eqnarray}
\sigma_{k}&\equiv&\frac{U}{\hbar}+2\mu_0-k\omega\\
\widetilde{\sigma}_{k}&\equiv&-\frac{U}{\hbar}+2\mu_0-k\omega\\
\widetilde{\widetilde{\sigma}}_{k}&\equiv& 2\mu_0-k\omega\;.
\end{eqnarray}
The model is still equivalent to the full Schr\"odinger equation corresponding to Eq.~(\ref{eq:H}); it now reads:
\begin{widetext}
\begin{equation}
\label{eq:DGLSa}
\hspace*{-2.5cm}
i\hbar\partial_t\left(\begin{array}{c}a_{0,0}\\a_{0,2}\\a_{2,0}\\a_{1,0}\\a_{1,1}\\a_{0,1}\end{array}\right)=\left(\begin{array}{cccccc} 0 & 0 & 0 & A_1(t)  & 0 & 0\\
                                                                                                           0 & 0 & 0 & 0 & A_2^*(t)& 0\\
                                                                                                         0 & 0 & 0 &  A_2^*(t) & A_1(t)  & 0\\
                                                                   A_1^*(t) & 0 &  A_2(t) & 0 & 0 &  A_3(t) \\
                                                                   0 &  A_2(t) &   A_1^*(t) & 0 & 0 &   A_3^{*}(t) \\
                                                                   0 & 0 & 0 &  A_3^{*}(t) &   A_3(t)  & 0 \\ \end{array}\right)
\left(\begin{array}{c}a_{0,0}\\a_{0,2}\\a_{2,0}\\a_{1,0}\\a_{1,1}\\a_{0,1}\end{array}\right)
\end{equation}
\end{widetext}
where
\begin{eqnarray}
\label{eq:vieleFrequenzen}
A_1(t)\equiv
-J\sqrt{2}\sum_{k}{\cal J}_{k}\left({{\textstyle\frac{2\mu_1}{\omega}}}\right)i^{k}
\exp({i\sigma_k t})\;,\\
A_2(t)\equiv
-J\sqrt{2}\sum_{k}{\cal J}_{k}\left({{\textstyle\frac{2\mu_1}{\omega}}}\right)i^{k}
\exp({i\widetilde{\sigma}_k t})\;,\\
A_3(t)\equiv
-J\sum_{k}{\cal J}_{k}\left({{\textstyle\frac{2\mu_1}{\omega}}}\right)i^{k}
\exp({i\widetilde{\widetilde{\sigma}}_k  t})\;.
\label{eq:vieleFrequenzen2}
\end{eqnarray}

 In general, for given parameters of tilt, interaction and shaking frequencies, there will be only one term with slow time-dependence~\footnote{{Cases for which the above sums involve two frequencies with opposite sign but similar magnitude are not likely to be covered by this approximation. Furthermore, for the slowly varying term to be really the dominant term, it has to be assumed that its prefactor is not much smaller than that of the other terms in the sum.}}. Following the reasoning of the rotating wave approximation~\cite{HarocheRaimond06}, this term should dominate the dynamics for not too large timescales. 
Defining 
\begin{equation}
\label{eq:simplified}
\sigma_{k'}\equiv\frac{U}{\hbar}+2\mu_0-k'\omega
\end{equation}
where $k'$ is the integer for which $|\frac{U}{\hbar}+2\mu_0-k\omega|$ reaches its minimum and repeating the same reasoning for $\widetilde{\sigma}_k$ and  $\widetilde{\widetilde{\sigma}}_k$ leads to a much simpler version of the matrix in Eq.~(\ref{eq:DGLSa}) which surprisingly can even be solved via exact diagonalisation~\footnote{{The time-dependent model~(\ref{eq:H}) can, in general, only be solved by numerically integrating the time-dependent Schr\"odinger equation. Within this paper, we do this by using the Shampine-Gordon routine~\cite{ShampineGordon75}; exact diagonalisation is a much simpler approach.}} (see Ref.~\cite{EsmannEtAl10IOP} where even some analytical results could be derived).
The derivation of our simplified model explains [cf.\ Eq.~(\ref{eq:simplified})] why the two curves in Fig.~\ref{fig:exactdiagonal} agree well except for the deviation for larger times.

\section{\label{sec:How}How to distinguish two-particle NOON-states from statistical mixtures}
The main idea is to be able to distinguish quantum superpositions from statistical mixtures via the behaviour concerning the return to the initial state. Why a tilted three-well potential is useful for considerations like that can already be motivated on the single-particle level:
On the one hand, despite the tilt the tunnelling processes
\begin{equation}
|0,1,0\rangle \to \frac1{\sqrt{2}}\left(|1,0,0\rangle\pm |0,0,1\rangle\right) \to |0,1,0\rangle
\end{equation}
do not involve any change of the potential energy. On the other hand, the individual tunnelling processes 
 \begin{equation}
\label{eq:stateinfach}
|0,1,0\rangle \to \left\{\begin{array}{c}|1,0,0\rangle\\ |0,0,1\rangle\end{array}\right\} \to |0,1,0\rangle.
\end{equation}
are related to a change in energy (because of the tilt) and thus both the first and the second tunnelling process of Eq.~(\ref{eq:stateinfach}) would thus be suppressed. This offers a starting point to distinguish the time-evolution of quantum superpositions from statistical mixtures: On the one hand, tunnelling of the quantum superposition  $\frac1{\sqrt{2}}\left(|1,0,0\rangle\pm |0,0,1\rangle\right)$ back to the initial state $|0,1,0\rangle$ is allowed energetically. On the other hand, for a statistical mixture like
\begin{equation}
\varrho = \frac12\left(|1,0,0\rangle\langle 1,0,0| + |0,0,1\rangle\langle 0,0,1|\right)
\end{equation}
the tunnelling to the state $|0,1,0\rangle$ is suppressed for energetic reasons.

There is, however, an even stronger reason why the return to the initial state can verify interesting quantum superpositions:

While experimentally measuring the fidelity~(\ref{eq:fidelity}) would either be impossible or at least be a considerable challenge for future experiments, measuring the number of atoms in one of the wells is comparatively straight-forward both in an optical superlattice~\cite{CheinetEtAl08} and in a single three-well potential~\cite{BakrEtAl09,ShersonEtAl10}. Thus, for the two-particle NOON-state $\frac1{\sqrt{2}}\left(|2,0,0\rangle+\exp(i\delta)|0,0,2\rangle\right)$
one could experimentally verify, that with nearly 50\% probability both particles are in the left well and with  50\% probability both particles are in the right well. In order to distinguish this situation from the statistical mixture,
\begin{equation}
\label{eq:corrmix}
\varrho = \frac12\left(|2,0,0\rangle\langle 2,0,0| + |0,0,2\rangle\langle 0,0,2|\right)\;,
\end{equation}
 we will be searching for parameters such that the probability to return to the initial state is high. As the initial state is the Fock state with two particles in the middle well, this is an unambiguous experimentally verifiable signature.

It remains to be shown that the return to the initial state at time $t_2$ -- in combination with the 50/50 probability to find both particles either in the left or in the right well at time $t_1<t_2$ -- really is unambiguous. Using the time-evolution operator $\hat{U}$, one has:
\begin{equation}
|\psi(t_2)\rangle = \hat{U}(t_2,t_1)|\psi(t_1)\rangle\;.
\end{equation}
Let us consider one particular ideal set of parameters for which the wave-function~$\psi_1(t)$ reaches the perfect cat-state~(\ref{eq:cat}) at $t=t_1$ followed by a probability to return to the initial state at $t=t_2$ of 100\%:
\begin{equation}
|\langle 0,2,0|\psi_1(t_2)\rangle|^2 =1\;.
\end{equation}
Within this ideal set of parameters, an inaccurate preparation of the wave-function  could 
lead to a return-probability of just $x$:
\begin{equation}
|\langle \psi_2(t_2)|\psi_1(t_2)\rangle|^2 =x\;,\quad 0\le x\le 1\;.
\end{equation}
The fact that $\hat{U}^{\dag}\hat{U}=1$ implies:
\begin{equation}
|\langle \psi_2(t_1)|\psi_1(t_1)\rangle|^2 =x\;.
\end{equation}
Thus, for our example of a perfect return to the initial state, a fidelity of $x$ at $t=t_1$ leads to a probability of $x$ to return to the initial state at $t=t_2$. In particular, if a NOON-state has a 100\% return probability to the initial state at a given time, the corresponding statistical mixture~(\ref{eq:corrmix}) will only manage a return to the initial state with 50\% probability at the same point of time~\footnote{{To see this, one can use the Schr\"odinger equation seperately for the time evolution of $|2,0,0\rangle$ and  $|0,0,2\rangle$.}}. While we will investigate one particular source of decoherence -- particle losses -- in the next section, other sources of decoherence would also lead to statistical mixtures and thus to a different dynamics. Given the level at which today's experiments with ultra-cold atoms are performed, we can assume that the particles do mostly behave as predicted by the Schr\"odinger equation without decoherence.

 If the return to the initial state is less than the expected 100\% (but still well above 50\%), this does not necessarily imply that the NOON-state was partially destroyed via decoherence: experimental uncertainties which would average over different solutions of the Schr\"odinger equation could produce a similar effect. Nevertheless, such a high return probability can still be a signature of the NOON-state: It has to be checked if the variations of the parameters still lead to high fidelities at $t=t_1$. In an experiment this would lead to probabilities of the order of 50/50 to find the two particles either in the left or in the right well.

\section{\label{sec:NOON}NOON-states with fidelities\\ above 90\%}
\begin{figure}
\begin{center}
\includegraphics[width = 1.0\linewidth]{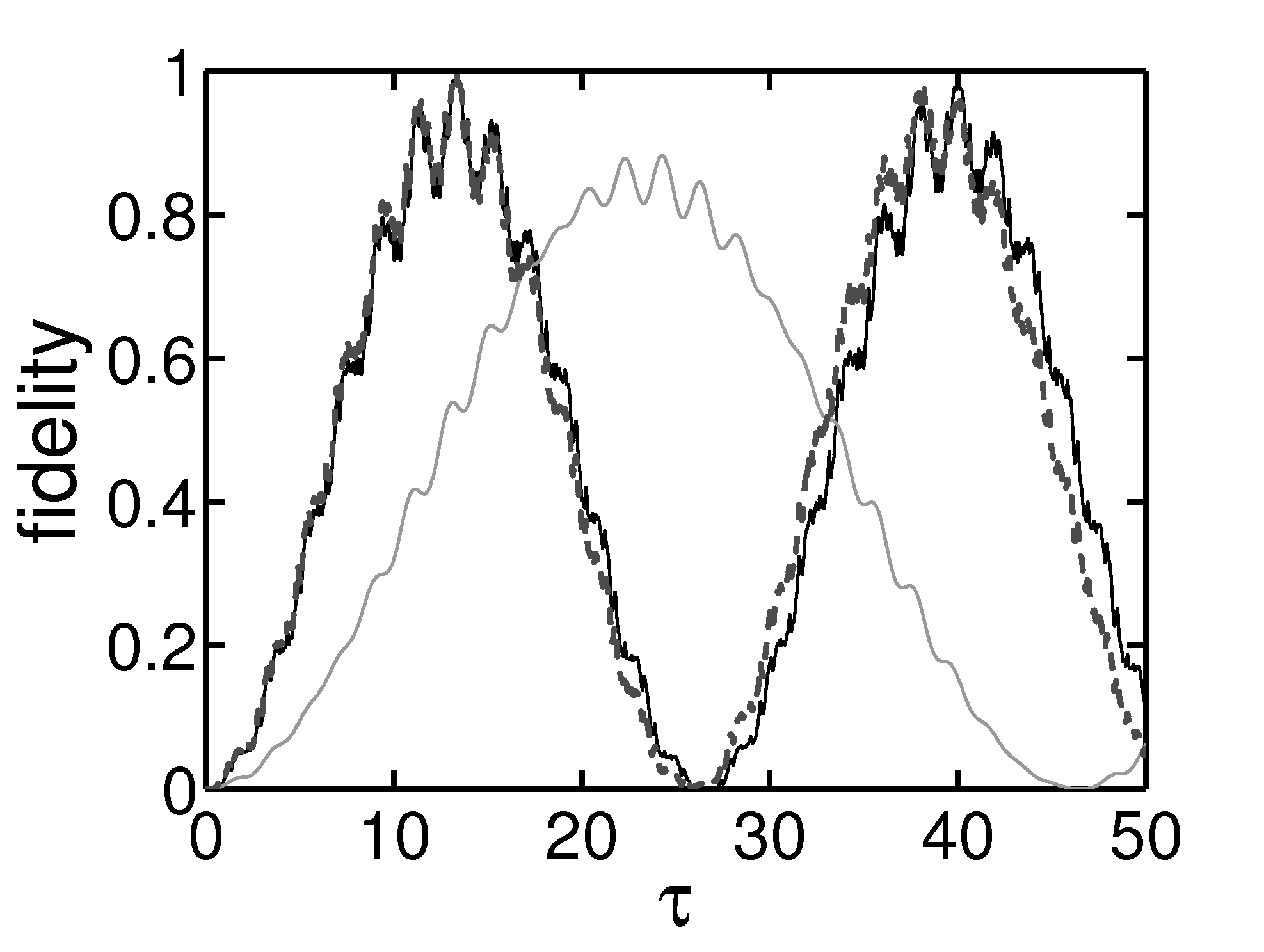}
\end{center}
\caption{\label{fig:fidgut}Fidelity~(\ref{eq:fidelity}) as a function of time for two particles initially sitting in the middle well and  $\hbar \omega=6J$, $U=9.1J$ in an untilted three-well potential ($\mu_0=0$) with a shaking amplitude of $\hbar \mu_1=18.3J$. Black solid line: numeric solution of the model~(\ref{eq:H}); the fidelity reaches values of 99.67\%,  well above 90\%. Gray solid line: the simplified model~(\ref{eq:simplified}) does not even predict the correct frequency of the oscillations; taking 15 terms in Eqs.~(\ref{eq:vieleFrequenzen})-(\ref{eq:vieleFrequenzen2}), again leads to a better agreement (dashed black line).}
\end{figure}

\begin{figure}
\begin{center}
\includegraphics[width = 0.9\linewidth]{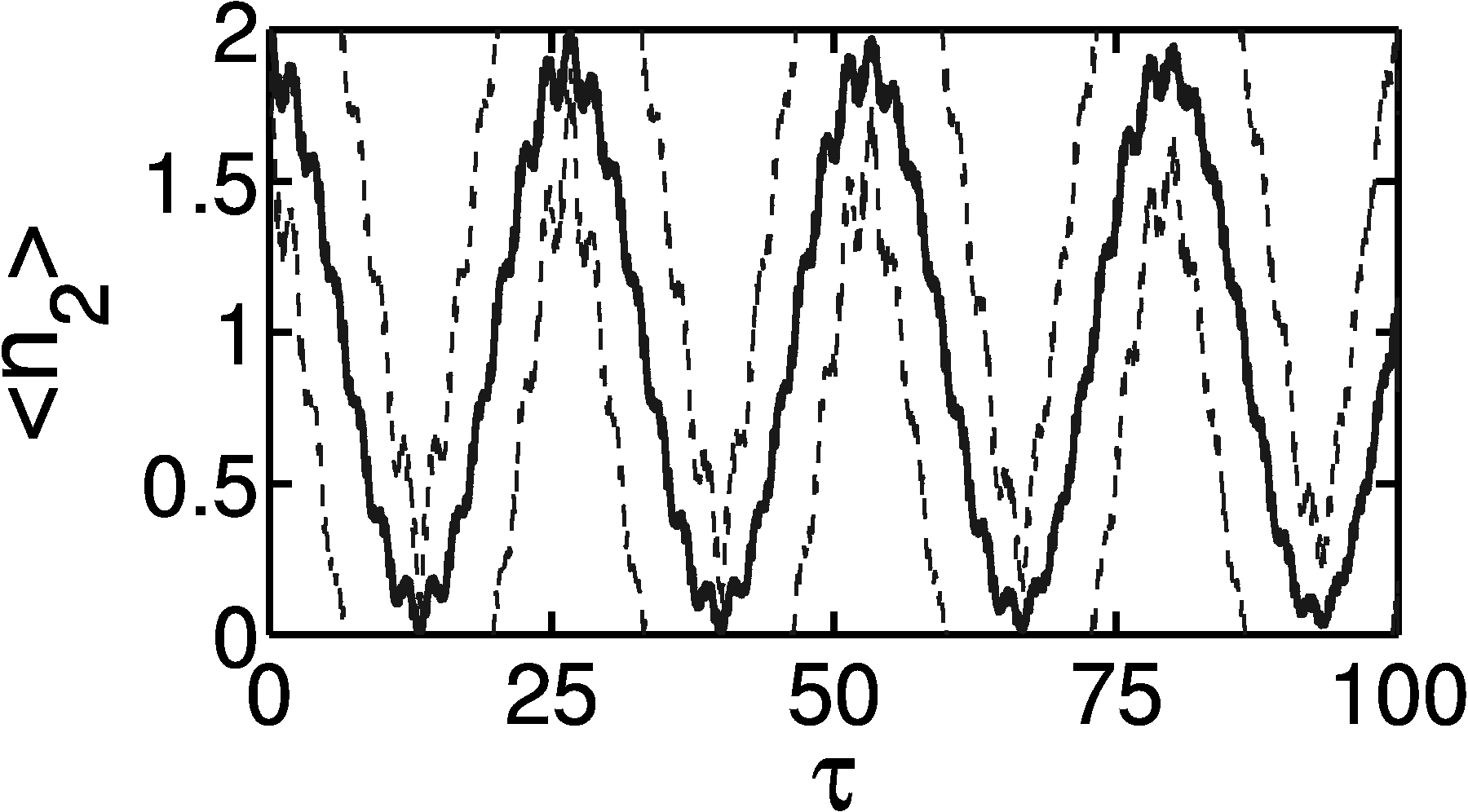}
\includegraphics[width = 0.9\linewidth]{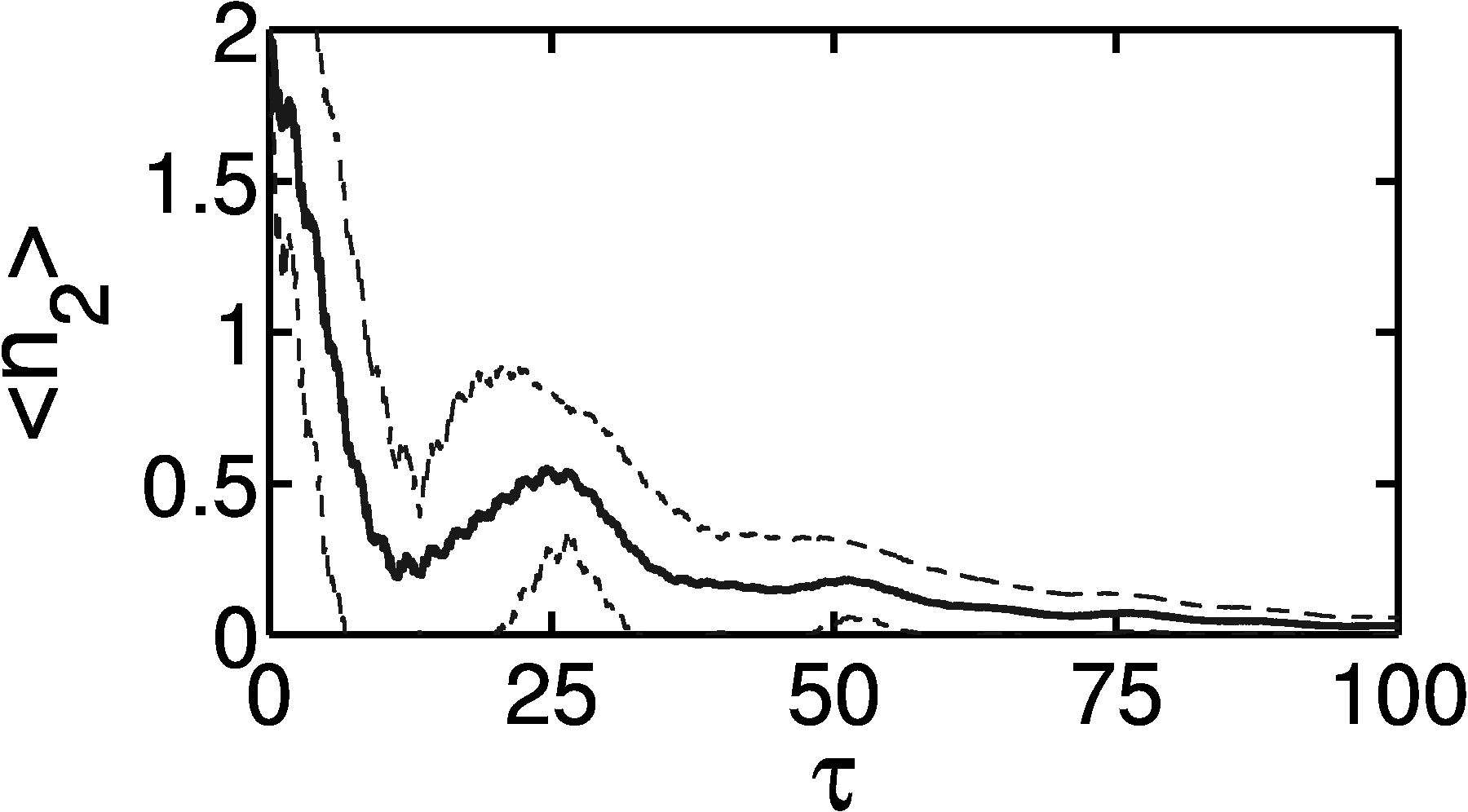}
\includegraphics[width = 0.9\linewidth]{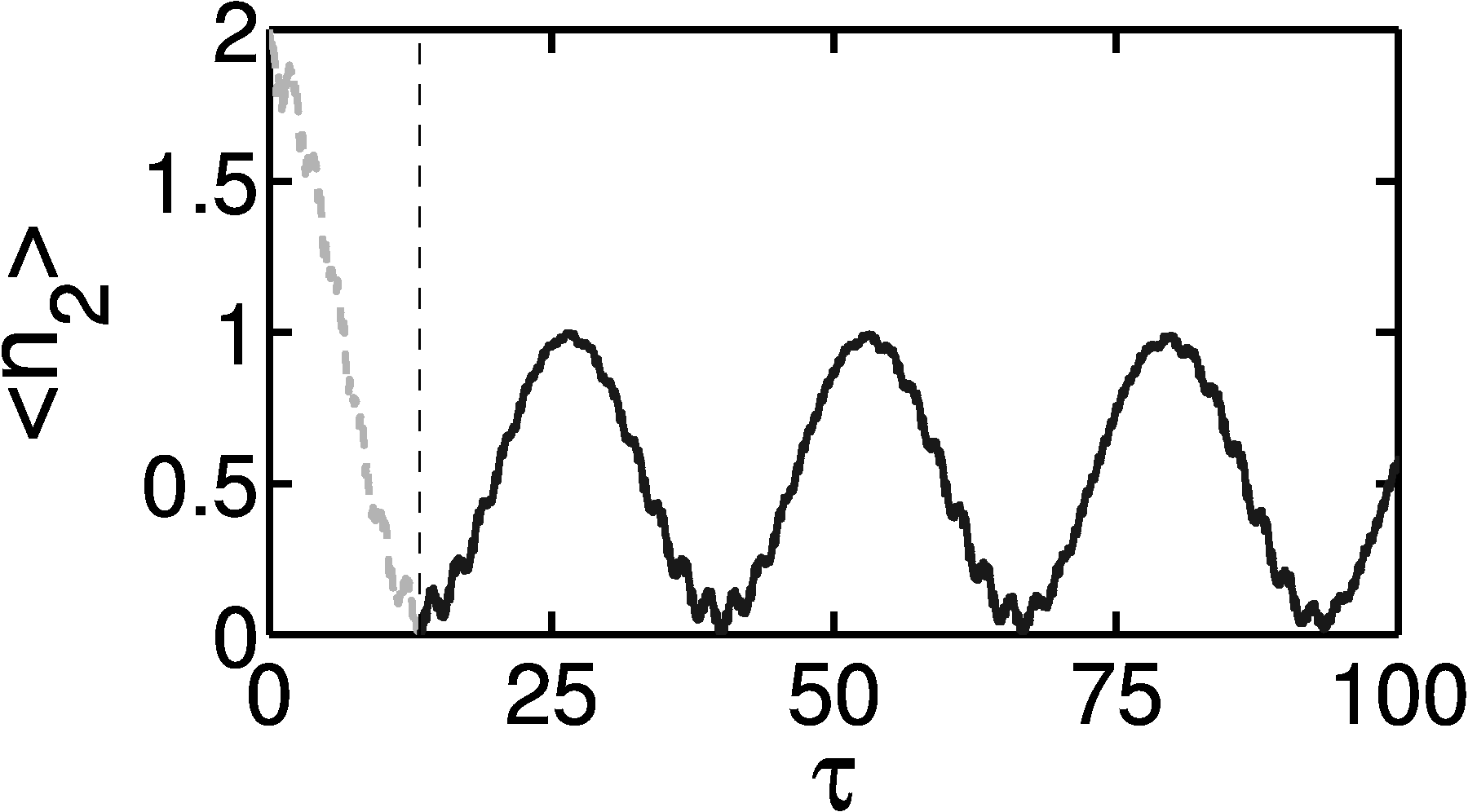}
\end{center}
\caption{\label{fig:dekohaerenz}Average number of particles in the second well as a function of time for the same parameters as in Fig.~\ref{fig:fidgut}. Upper panel:  thick solid line: $\langle n_2\rangle$ as predicted by quantum dynamics obtained by numerically solving the model~(\ref{eq:H}); at $\tau \simeq 13.36$ there are hardly any particles in the middle well ($\langle n_2\rangle = 0.026$) and at $\tau \simeq 26.69$ the particles have returned to the middle well ($\langle n_2\rangle =1.9919$).  Dashed lines:   $\langle n_2\rangle\pm \Delta  n_2$ where  $\Delta  n_2$ is the standard deviation. Middle panel: thick solid line: $\langle n_2\rangle$ as predicted by a combination of quantum dynamics and particle losses~(\ref{eq:nvont}) for $\alpha=J/(30\hbar)$. Black dashed lines:  $\langle n_2\rangle\pm \Delta  n_2$ where  $\Delta  n_2$ is the standard deviation. Lower panel:  $\langle n_2\rangle$ as predicted by quantum dynamics obtained by numerically solving the model~(\ref{eq:H}) if the quantum superposition turns into a statistical mixture at the point of highest fidelity ($\tau_{\rm max}=13.356$). The required accuracy of the parameters is discussed above Eq.~(\ref{eq:btttt}).
}
\end{figure}

To obtain better results than displayed in Figs.~\ref{fig:exactdiagonal} and \ref{fig:ohnetreiben}, three requirements were used for the numeric optimisation of the parameters:
\begin{itemize}
\item a fidelity well above 90\%.
\item a high probability to return to the initial state.
\item not too large timescales.
\end{itemize}
There are parameter-sets with and without tilt that  fulfil these conditions. One example is shown in Figs.~\ref{fig:fidgut} and \ref{fig:dekohaerenz}. 
The approach via the simplified model~(\ref{eq:simplified}) only works if more than one frequency is included in  Eqs.~(\ref{eq:vieleFrequenzen})-(\ref{eq:vieleFrequenzen2}) (see Fig.~\ref{fig:fidgut}). At first glance, the standard deviation displayed in Fig.~\ref{fig:dekohaerenz} might render it difficult to distinguish the behaviour including decoherence from the ideal quantum case. However, for the accuracy with which the average can be measured, the standard error is relevant which is smaller by a factor of~$1/\sqrt{M}$ for $M$ measurements. Taking $M\simeq 100$ this would thus reduce the error by a factor of 10.  
Our theory so far ignores decoherence. For the present system, atom losses via scattering with back-ground atoms are an important decoherence mechanism (cf.\ Refs.~\cite{WeissCastin09,KohlerSols02}). To include this into our model, we use a method called \textit{piecewise deterministic processes}, PDP in Ref.~\cite{Breuer06}, which is based on Refs~\cite{DalibardEtAl92,DumEtAl92}. The main idea is to use quantum dynamics for the time-evolution in combination with atom losses for which the points of time of the atom losses are chosen via random numbers~\cite{Breuer06,LiEtAl08}. The results displayed in the middle panel of Fig.~\ref{fig:dekohaerenz} are averaged over many sets of random numbers. Atoms are lost with a loss-rate  $\alpha$ (cf.\ \ref{sec:Lindblad}), 
\begin{equation}
\label{eq:nvont}
\frac{d}{dt}\langle N(t)\rangle = -\alpha \langle N(t)\rangle\;,
\end{equation}
corresponding to an exponential decay of the average number of particles, which for the two particles reads:
\begin{equation}
\label{eq:nmittel}
\langle N(t)\rangle =2\exp(-\alpha t)\;.
\end{equation}
The middle panel of Fig.~\ref{fig:dekohaerenz} uses  $\alpha=\alpha_1$ with $\alpha_1=J/(30\hbar)$; for the timescales of the double-well lattice experiment~\cite{CheinetEtAl08}, the fidelity reaches its maximum at $9.7\ldots 16.1\,$ms and thus $\alpha\approx 1/36\ldots 1/22\,$kHz. In order to be able to observe the return to the initial state with 1\% accuracy, we require the average number of particles~(\ref{eq:nmittel}) to be at least 99\% of the initial value up to a timescale for which pure quantum evolution predicts a nearly perfect return to the initial state ($32\,$ms). This results in a lifetime of the particles of~$1/\alpha_2\approx 3\,$s. This realistic lifetime of a single particle in the lattice lies two orders of magnitude below the lifetime required to produce NOON-states of, say, 100 particles according to the scheme of Ref.~\cite{WeissCastin09}. 

The lower panel of Fig.~\ref{fig:dekohaerenz} shows what would happen if the quantum superposition was turned into a statistical mixture at the time for which the fidelity reaches its maximum: although now there still are two particles involved in the simulation, the average occupancy of the middle well would remain a factor of two below the values reached for pure quantum evolution. 

\begin{figure}
\includegraphics[width = \linewidth]{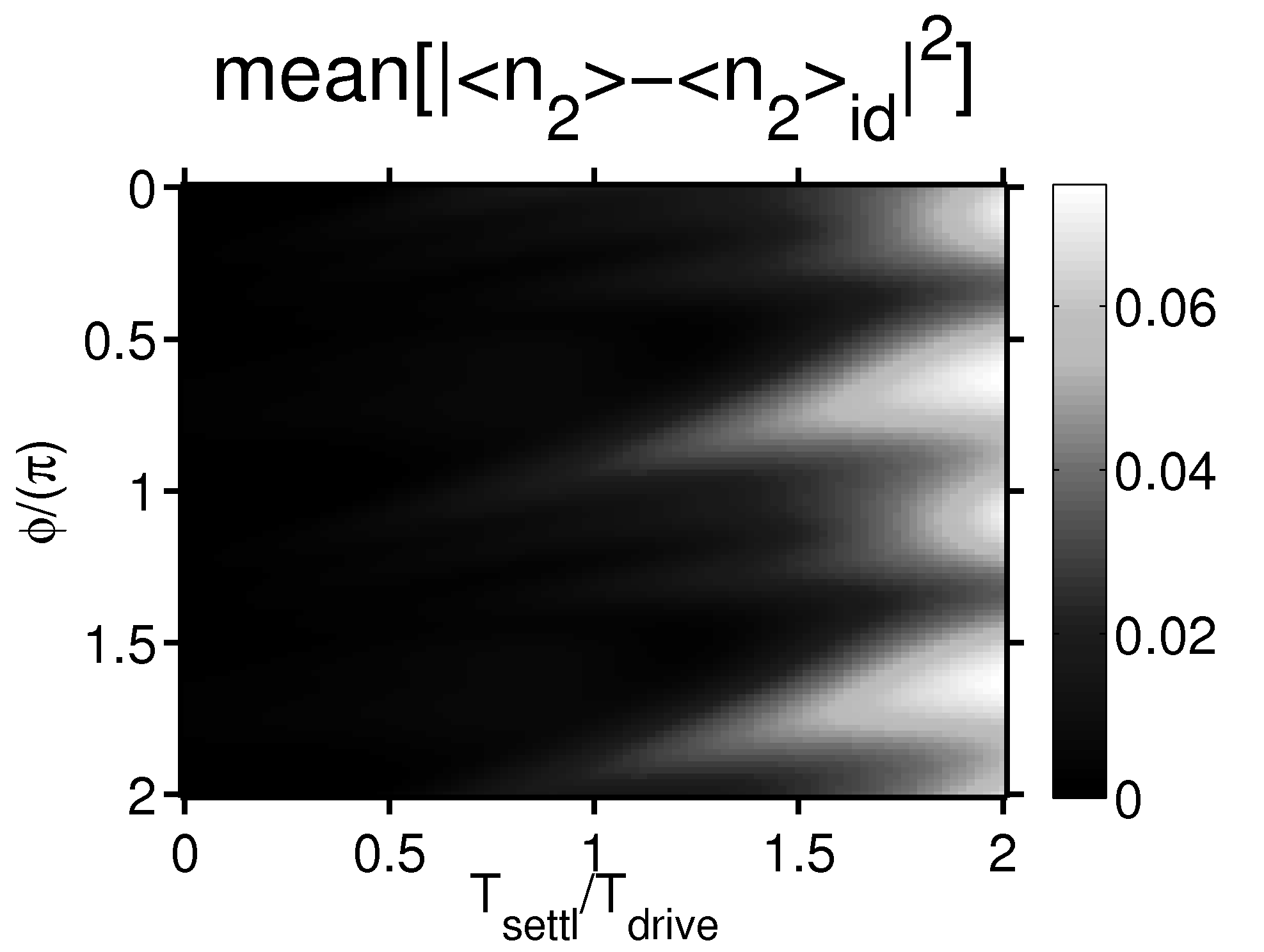}
\caption{\label{fig:mean}Mean square deviation  (\ref{eq:mean}) for $T = 100\hbar/J$ in two-dimensional projection as a function of both settling time $T_{\rm settl}$ [see Eq.~(\ref{eq:settl})]  and initial phase.  All other parameters as in Fig~\ref{fig:fidgut}. $T_{\rm drive}\equiv 2\pi/\omega$ is the oscillation period. Contrary to experiments like~\cite{HallerEtAl10}, in our case the initial phase does not play any role.
}
\end{figure}
In order to reproduce the generation of NOON-states shown in Figs.~\ref{fig:fidgut} and \ref{fig:dekohaerenz}, an experiment would require a control of both interaction ($U/J$) and driving amplitude $\hbar\mu_1$ of the order of several per cent, the tilt~$2\hbar\mu_0$ should be zero to about $0.04 J/\hbar$ in order to reach fidelities above 0.95~\footnote{{The requirements on the accuracy of the parameters were obtained by calculating both the fidelity and the return to the initial states at the time defined by the ideal case (rather than using the maximum fidelity/maximum return for each parameter set).}}.
In an experiment, there might be problems in realising a periodic shaking which is exactly zero for times $t<0$ and perfectly sinusoidal for times $t>0$. To show that this is not necessarily a problem, Fig.~\ref{fig:mean} shows what happens if the shaking amplitude is switched on and off with $T_{\rm settl}$ as the characteristic timescale:
\begin{eqnarray}
\label{eq:btttt}
 B(t)&=&2\hbar\mu_1\sin(\omega t+\phi)\left[1-\exp\left({\textstyle-\frac{2\pi t}{T_{\rm settl}}}\right)\right]\nonumber\\
 &\times&
\left[1-\exp\left({\textstyle-\frac{2\pi(T-t)}{T_{\rm settl}}}\right)\right]\label{eq:settl}\;,\nonumber\\
 &&0\le t\le T\,;
\end{eqnarray}
 the shaking now includes a phase $\phi$ in $\sin(\omega t +\phi)$. Plotted is the mean deviation from the case without any increase of the amplitude:
\begin{equation}
\label{eq:mean}
{\rm mean}\left[\left|\langle n_2\rangle -\langle n_2\rangle_{\rm id}\right|^2\right] \equiv \frac 1T\int_0^T 
\left|\langle n_2\rangle -\langle n_2\rangle_{\rm id}\right|^2 dt
\end{equation}
where $\langle n_2\rangle_{\rm id}$ is the model~(\ref{eq:H}).

\section{Conclusion}

To conclude, we have shown via numerical investigations that high-fidelity-two-particle NOON-states can emerge in three-well potentials. 
There is, in particular, an easily observable difference between statistical mixtures and perfect NOON-states: only the latter would return to the initial states. For realistic conditions, decoherence would be present in such experiments, which would destroy the quantum superpositions. Here, the proposed scheme might offer the possibility to measure the timescales of decoherence. 

Three-well lattices could be used to realise such experiments - produced either via a superposition of lasers similar to the double-well lattices of Refs.~\cite{CheinetEtAl08,Folling07} or by using subwavelength lattices~\cite{YiEtAl08}. The number of atoms per three-well lattice can be controlled precisely via the Mott-insulator used to load the lattice~\cite{CheinetEtAl08}. Such a setup would allow to measure the average number of particles of one of the three lattice sites in a single experiment by averaging over all three-well potentials. Furthermore, by focusing on one single three-well potential within the three-well lattice via techniques similar to  Refs.~\cite{BakrEtAl09,ShersonEtAl10} and repeating the experiments many times, even more precise measurements could be possible.

\section*{Acknowledgements}

We thank M.\ Holthaus for his support and H.-P.\ Breuer, M.\ Esmann and F.\ Schreck  for discussions. 
 BG acknowledges funding by
the Studienstiftung des deutschen Volkes. NT acknowledges funding by the University of Oldenburg. CW acknowledges support     by C.\ \& C.\ Hettich in initiating this project.\nocite{ShampineGordon75}

\begin{appendix}
\section{\label{sec:Lindblad}Lindblad equation to model particle losses}

A derivation of the Lindblad Master equation can be found, e.g., in Ref.~\cite{Breuer06}. Often, $a^{(\dag)}$ is the annihilation (creation) operator of a photon in mode~$\omega$ but atom losses can be described by the same approach: 
\begin{eqnarray}
\label{eq:lindblad}
\frac{d}{dt}\hat{\varrho} =&&
-i\omega [a^{\dag}a,\hat{\varrho}] -\frac{\kappa_-}2\left(a^{\dag}a\hat{\varrho}+\hat{\varrho} a^{\dag}a - 2a\hat{\varrho} a^{\dag}\right)\nonumber\\
&&-\frac{\kappa_+}2\left(aa^{\dag}\hat{\varrho}+\hat{\varrho}a a^{\dag} - 2a^{\dag}\hat{\varrho} a\right)\;,
\end{eqnarray}
where $\hat{\varrho}$ is the density matrix.
We assume that particles only are lost, thus $\kappa_+=0$ and $\alpha\equiv \kappa_-$ and thus derive a rate equation~\cite{HarocheRaimond06} by using the fact that the photon number distribution $p_n(t)$ is related to the density matrix via
\begin{equation}
p_n(t) = \langle n|\hat{\varrho}|n\rangle \;.
\end{equation}

The rate-equation thus reads:
\begin{eqnarray}
\label{eq:masterp}
\frac {d}{dt}p_n(t) &=& \alpha(n+1)p_{n+1}(t) 
-\alpha np_n(t)\;.
\end{eqnarray}
Using this rate-equation, it is straight-forward~\cite{HarocheRaimond06} to derive the exponential decay of the numbers of particles given in Eq.~(\ref{eq:nvont}). As explained at the beginning of Sec.~\ref{sec:NOON}, we use piecewise deterministic processes~\cite{Breuer06} (cf.~\cite{DalibardEtAl92,DumEtAl92})
 to model the atom losses; in between loss events, the Hamiltonian dynamics of Eq.~(\ref{eq:H}) is used. Further theoretical investigations of atom losses can be found, e.g., in Ref.~\cite{LiEtAl08}.
\end{appendix}


%

\end{document}